\documentclass[final,times,5p,sort&compress]{elsarticle}

\usepackage{lineno,hyperref}

\usepackage{amsmath}
\usepackage{amssymb}
\usepackage{graphicx}
\usepackage{bm}
\usepackage{braket}
\usepackage{mathrsfs}
\usepackage{accents}
\usepackage{longtable}
\usepackage{rotating}
\usepackage{cancel}

\modulolinenumbers[5]

\journal{Physics Letters B}

\begin{document}

\begin{frontmatter}

\title{
Spin entanglement in neutron-proton scattering
}

\cortext[mycorrespondingauthor]{Corresponding author}

\author[hhu]{Dong Bai \corref{mycorrespondingauthor}}
\ead{dbai@hhu.edu.cn}

\address[hhu]{College of Science, Hohai University, Nanjing 211100, China}

\begin{abstract}

In this Letter, I work out spin
entanglement properties of neutron-proton scattering using the \emph{exact} $S$-matrix,
generalizing previous works based on $S$ wave. 
The dependence of spin entanglement on momentum, scattering angle, and initial spin configuration is investigated for realistic nuclear forces,
while low-energy properties of spin entanglement are analyzed 
within the framework of pionless effective field theory at leading order.
New connections are found between spin entanglement and symmetry enhancement of strong interactions.
These results
lead to a more complete understanding of how spin entanglement is generated via neutron-proton interaction. 
They also lay the theoretical foundation for controllable production of entangled nucleon-nucleon pairs in future experiments.

\end{abstract}


\date{}%

\end{frontmatter}


\section{Introduction}

The meaning of quantumness has kept evolving since Planc- k's original proposal of the quantum hypothesis in 1900.
Nowadays, it may have three related yet different aspects.
The first aspect may be summed up by the phrase ``discrete energies'',
which refers to the fact that 
 energies of microscopic systems can take discrete values \cite{Landau:1981,Messiah:2014,Weinberg:2015,Sakurai:2020}.
It plays a key role in explaining the black-body radiation and the stability of the hydrogen atom.
The second aspect of quantumness may be summed up by the phrase ``quantum ripples of fields'',
which refers to the field-theory origin of quantum particles \cite{Weinberg:1995,Peskin:1995,Schwartz:2013,Coleman:2018}.
According to quantum field theories (QFTs),
the spacetime is filled with different kinds of fields,
with the vacuum identified as their ground state
and quantum particles as their quanta, i.e., quantized excitations.
This is the foundation of the standard model in high energy physics
and is also important in nonrelativistic quantum many-body physics.
The third aspect of quantumness may be summed up by the phrase ``spooky action at a distance'',
which refers to entanglement, a special kind of correlation in multipartite quantum systems that has no classical analogy \cite{Horodecki:2009zz,Nielsen:2010,Witten:2018zva,Benenti:2019}.
Technically speaking, entanglement refers to the phenomenon where the quantum state of a multipartite system 
cannot always be expressed as a tensor product of individual quantum states of its constituents.
It is at the core of many quantum algorithms and quantum communication protocols in quantum information science,
marking it a hallmark of the second quantum revolution.

The nucleus is a composite system made up of nucleons.
Understanding its quantum properties is the primary focus of nuclear physics.
The field has made notable theoretical achievements that can be classified based on different aspects of quantumness.
For instance, the nuclear shell model, which is the most popular microscopic model in nuclear physics, 
shows how nucleons occupy discrete energy levels supported by the nuclear mean field potential with a large spin-orbit term \cite{Mayer:1957}. 
This highlights the first aspect of nuclear quantumness.
In Yukawa's meson theory, which paved the way to the modern theory of nuclear forces, 
it is suggested that the nuclear force between two nucleons is mediated by a field whose quantum is known as meson, 
much like the electric force between two charged particles, which is mediated by the electromagnetic field with the photon as its quantum \cite{Yukawa:1935}.
This demonstrates the second aspect of nuclear quantumness. 
But what about the third aspect of nuclear quantumness? 
What is the most significant theoretical achievement concerning the entanglement properties of the nucleus?

When compared to the first two aspects,
it is fair to say that the third aspect of nuclear quantumness, the quantum entanglement aspect of the nucleus, is less explored. 
Only recently has this topic begun to attract some attention \cite{Kwasniewicz:2013cqa,Kanada-Enyo:2015ncq,Kanada-Enyo:2015kyo,Legeza:2015fja,Gorton:2018,Johnson:2019,Kruppa:2020rfa,Kovacs:2021yme,Kwasniewicz:2016,Kwasniewicz:2017dbc,Beane:2018oxh,Robin:2020aeh,Low:2021ufv,Beane:2021zvo,Jafarizadeh:2022kcq,Pazy:2022mmg,Tichai:2022bxr,Bai:2022hfv,Liu:2022grf,Johnson:2022mzk,Bai:2023rkc,Gu:2023aoc}.
Despite several interesting findings, many crucial questions remain unanswered.
Some of the most pressing questions include:
What kinds of entanglement properties does the nucleus have?
What are the relations between nuclear entanglement and nuclear forces?
And how can the study of entanglement structures help us simulate nuclear dynamics on classical and quantum computers? 
Pursuing the answers to these questions would undoubtedly bring numerous benefits to the field of nuclear physics.

Nucleon-nucleon scattering is a fundamental process in low-energy nuclear physics.
On-shell nucleon-nucleon scattering provides key inputs to constrain theoretical models of nuclear forces \cite{Epelbaum:2008ga,Machleidt:2011zz,Hammer:2019poc,Epelbaum:2019kcf}, 
while off-shell nucleon-nucleon scattering is essential to some microscopic approaches to nucleon-nucleus and nucleus-nucleus scattering \cite{Hebborn:2022vzm}.
In order to understand the entanglement properties of the nucleus,
it is crucial to first understand the entanglement properties of nucleon-nucleon scattering comprehensively.
It will lay the foundation for understanding the emergence of entanglement structures in the nucleus
from the microscopic viewpoint.
 
Very recently, entanglement properties of neutron-proton scattering have been pioneered by several authors in spin space \cite{Beane:2018oxh,Low:2021ufv,Bai:2022hfv,Liu:2022grf,Bai:2023rkc}.
Their analyses are focused on the $S$-wave component of neutron-proton $S$-matrix,
which has the advantage of simple structures and leads to an analytic understanding of spin entanglement.
In this Letter, I generalize these studies by working out spin entanglement properties of \emph{exact} neutron-proton $S$-matrix, rather than just its $S$-wave component.
Numerical results are discussed in depth from different perspectives, which may shed light on controllable production of entangled nucleon-nucleon pairs in future experiments.
This Letter is structured as follows: 
In Sec.\ II, some of the most important properties of neutron-proton $S$-matrix are reviewed first, followed by a brief introduction to the entanglement power and the concurrence,
two measures of spin entanglement.
In Sec.\ III, the properties of spin entanglement are analyzed carefully, along with its relation to symmetry enhancement of strong interactions at low energies. 
Finally, Section IV concludes and summarizes the main findings.


\section{Theoretical formalism}

\label{SA}

\subsection{Neutron-proton S-matrix}

 \begin{figure}

\centering
  \includegraphics[width=0.75\linewidth]{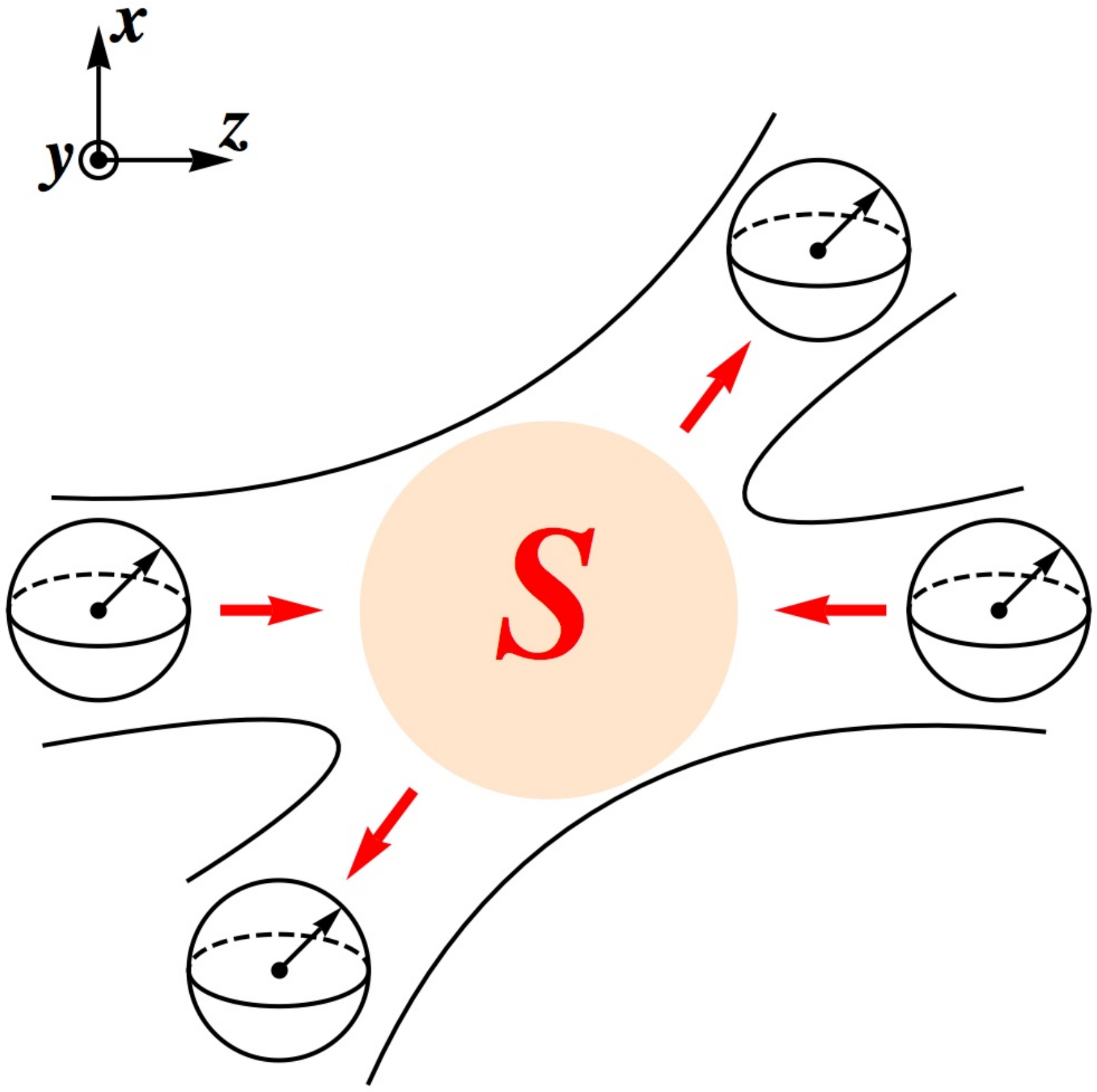}
  \caption{
An illustration of neutron-proton scattering, regarding nucleons as qubits.
Bloch spheres are adopted to represent qubits in a geometric way.
See the text for details.
}

  \label{QS}
       
\end{figure}

Proton and neutron are spin-$1/2$ particles. They can be regarded safely as qubits below thresholds of $\Delta$ excitations, 
with spin-up and spin-down states identified with $\ket{\uparrow}=\ket{0}=[1,0]^T$ and $\ket{\downarrow}=\ket{1}=[0,1]^T$.
An illustration of neutron-proton scattering in the center-of-mass frame is shown in Fig.~\ref{QS},
where nucleons are depicted by Bloch spheres, 
the standard geometric representation of qubits in quantum information science.
Initially, the two nucleons get prepared separately in the distant past, sharing no entanglement.
Then, they fly towards each other along the horizontal direction
to interact and get entangled inside the interaction domain (the light orange disk in the center of Fig.~\ref{QS}). 
Finally, they fly apart along another direction and move into the distant future,
where detectors can be arranged to measure the spin states of outgoing nucleons.

The neutron-proton $S$-matrix element $S_{m_1'm_2'm_1m_2}(\bm{p}',\bm{p})= \braket{\bm{p}'m_1'm_2'|S|\bm{p}\,m_1m_2}$ can be parametrized as follows:
\begin{align}
S_{m_1'm_2'm_1m_2}(\bm{p}',\bm{p})
=\frac{\delta(E_{p'}-E_p)}{\mu p}\hat{S}_{m_1'm_2'm_1m_2}(\bm{p}',\bm{p}),
\end{align}
where $\bm{p}$ and $\bm{p}'$ are relative momenta of in and out states, 
$E_p={p^2}/{(2\mu)}$ and $E_{p'}={p'^2}/{(2\mu)}$ are relative energies, 
$\mu$ is the two-body reduced mass,
and $m_1$, $m_2$, $m_1'$, $m_2'$ $=\pm\frac{1}{2}$
are spin indices of neutron and proton.
For the reduced $S$-matrix element $\hat{S}_{m_1'm_2'm_1m_2}(\bm{p}',\bm{p})$,
the on-shell condition $|\bm{p}'|=|\bm{p}|$ is always imposed.
As $S$-matrix is a unitary operator,
$\hat{S}_{m_1'm_2'm_1m_2}(\bm{p}',\bm{p})$ satisfies
\begin{align}
&\sum_{m_1''m_2''}\int\mathrm{d}^2\hat{p}''\hat{S}^\dagger_{m_1'm_2'm_1''m_2''}(\bm{p}',\bm{p}'')\hat{S}_{m_1''m_2''m_1m_2}(\bm{p}'',\bm{p})\nonumber\\
=&\sum_{m_1''m_2''}\int\mathrm{d}^2\hat{p}''\hat{S}_{m_1'm_2'm_1''m_2''}(\bm{p}',\bm{p}'')\hat{S}^\dagger_{m_1''m_2''m_1m_2}(\bm{p}'',\bm{p})\nonumber\\
=&\,\delta_{m_1'm_1}\delta_{m_2'm_2}\delta^2(\hat{p}'-\hat{p}),
\label{shatunitary}
\end{align}
%
with $\hat{p}=\bm{p}/p$ 
and $\hat{p}'=\bm{p}'/p'$ 
being the unit vectors along $\bm{p}$ and $\bm{p}'$.

Given the in state $\ket{\text{in}}=\ket{\bm{p}}\ket{\chi_\text{in}}$, with $\ket{\bm{p}}$ and $\ket{\chi_\text{in}}$ being the momentum and spin states\footnote{$\ket{\chi_\text{in}}$ is always assumed to be separable and \emph{pure}, corresponding to completely polarized neutron-proton scattering.
See Ref.~\cite{Bai:2023rkc} for quantum information properties of partially polarized neutron-proton scattering.}, 
the out state $\ket{\text{out}}= S\ket{\text{in}}$ is found to be
\begin{align}
&\ket{\text{out}}
=\int\mathrm{d}^2\hat{p}'\ket{\bm{p}'}\hat{\bm{S}}(\bm{p}',\bm{p})\ket{{\chi}_\text{in}},\label{out_mat}\\
&\hat{\bm{S}}(\bm{p}',\bm{p})=\sum_{\substack{m_1'm_2'\\m_1m_2}}\hat{S}_{m_1'm_2'm_1m_2}(\bm{p}',\bm{p})\ket{m_1'm_2'}\!\bra{m_1m_2}\label{shat_mat}\\
&\qquad\quad=\delta^2(\hat{p}'-\hat{p})\bm{1}_4+\frac{ip}{2\pi}\bm{M}(\bm{p}',\bm{p}),\label{spin_amp}
\end{align}
where $\bm{M}(\bm{p}',\bm{p})$, the spin amplitude, is a $4\times4$ matrix depending on $\bm{p}$ and $\bm{p}'$,
with rows and columns labeled by $(m_1,m_1)=(\frac{1}{2},\frac{1}{2}), (\frac{1}{2},-\frac{1}{2}), (-\frac{1}{2},\frac{1}{2}), (-\frac{1}{2},-\frac{1}{2})$ in order.
From Eqs.~\eqref{out_mat}--\eqref{spin_amp}, it is straightforward to see that, at a specific $\bm{p}'$ ($\neq\bm{p}$, the nonforward direction),
the out spin state is given by
\begin{align}
\ket{\chi_\text{out}}=\frac{\bm{M}(\bm{p}',\bm{p})\ket{{\chi}_\text{in}}}{\sqrt{\braket{\chi_\text{in}|\bm{M}^\dagger(\bm{p}',\bm{p})\bm{M}(\bm{p}',\bm{p})|\chi_\text{in}}}}\label{chi_out},
\end{align}
and the first term in Eq.~\eqref{spin_amp} makes no contribution to $\ket{\chi_\text{out}}$ in the nonforward direction.
As a $4\times4$ matrix, the spin amplitude $\bm{M}(\bm{p}',\bm{p})$ can be decomposed in terms of $\{\sigma_\mu\otimes\sigma_\nu\}$,
with $\sigma_\mu=(\bm{1}_2,\bm{\sigma})$ and $\bm{\sigma}=(\sigma_x,\sigma_y,\sigma_z)$.
Such a decomposition was first done by Wolfenstein in the 1950s \cite{Wolfenstein:1952}.
Later on, alternative amplitude systems were proposed by various groups, such as the Saclay \cite{Bystricky:1978,France:1980}, Hoshizaki \cite{Hoshizaki:1968}, helicity \cite{Jacob:1959}, and singlet-triplet systems \cite{Hoshizaki:1968}.
Explicitly, in the Saclay amplitude system, one has
\begin{align}
&\bm{M}(\bm{p}',\bm{p})\nonumber\\
=&\frac{1}{2}\bigg\{(a+b)+(a-b)(\bm{\sigma}_1\cdot\bm{n})(\bm{\sigma}_2\cdot\bm{n})+(c+d)(\bm{\sigma}_1\cdot\bm{m})(\bm{\sigma}_2\cdot\bm{m})\nonumber\\
+&\,(c-d)(\bm{\sigma}_1\cdot\bm{l})(\bm{\sigma}_2\cdot\bm{l})+(e+f)\bm{\sigma}_1\cdot\bm{n}+(e-f)\bm{\sigma}_2\cdot\bm{n}\bigg\},\label{Saclay}
\end{align}
where
$\bm{l}$, $\bm{m}$, $\bm{n}$ are the unit vectors given by
\begin{align}
\bm{l}=\frac{\bm{p}'+\bm{p}}{|\bm{p}'+\bm{p}|},
\quad
\bm{m}=\frac{\bm{p}'-\bm{p}}{|\bm{p}'-\bm{p}|},
\quad
\bm{n}=\frac{\bm{p}\times\bm{p}'}{|\bm{p}\times\bm{p}'|},
\end{align}
and $a$, $b$, $c$, $d$, $e$, $f$
are the six Saclay amplitudes, which are functions of the relative momentum $p$
and the relative angle $\theta$ between $\hat{p}$ and $\hat{p}'$.
Alternatively, $\bm{M}(\bm{p}',\bm{p})$ could be expressed in terms of phase shifts,
with the matrix element $M_{m_1'm_2'm_1m_2}(\bm{p}',\bm{p})
=\braket{m_1'm_2'|\bm{M}(\bm{p}',\bm{p})|m_1m_2}$ given by
\begin{align}
&M_{m_1'm_2'm_1m_2}(\bm{p}',\bm{p})\nonumber\\
=&\sum_{s}\braket{\tfrac{1}{2}m_1',\!\tfrac{1}{2}m_2'|s(m_1'+m_2')}\braket{\tfrac{1}{2}m_1,\!\tfrac{1}{2}m_2|s(m_1+m_2)}\nonumber\\
\times&\,{M}^s_{m_1'+m_2',m_1+m_2}(\bm{p}',\bm{p}),\label{FS}\\
&M^s_{m'm}(\bm{p}',\bm{p})\nonumber\\
=&\sum_{\substack{JL'LM}}i^{-L'+L}Y_{LM}^*(\hat{p})\braket{sm',L'(m+M-m')|J(m+M)}\nonumber\\
\times&\,Y_{L'(m+M-m')}(\hat{p}')\braket{sm,LM|J(m+M)}\left[{S}^J_{L'sLs}(p)-\delta_{L'L}\right],\label{Msmpm}
\end{align}
where $\braket{\tfrac{1}{2}m_1,\!\tfrac{1}{2}m_2|sm}$ is the Clebsch-Gordan coefficient for spin-$\frac{1}{2}$ particles,
with $s=0,1$ and $m=-1,0,1$,
and
the partial-wave $S$-matrix element $S^J_{L'sLs}(p)$ is related to phase shifts and mixing parameters as shown by Ref.~\cite{Gloeckle:1983}.

\subsection{Entanglement power}

The entanglement power is a physical measure that quantifies the entanglement generation capacity of a specific quantum operator \cite{Zanardi:2001}. It averages 
the so-called 2-entropy over all the possible unentangled in state
and is defined by
\begin{align}
\mathcal{E}(p,\theta)= 1-\int\frac{\mathrm{d}\Omega_1}{4\pi}\frac{\mathrm{d}\Omega_2}{4\pi}\text{tr}_1(\rho_1^2),\label{EP}
\end{align}
with $\rho_1$ being the reduced density matrix of $\ket{\chi_\text{out}}$ by tracing out the proton sector
\begin{align}
\rho_1&=\text{tr}_2\!\left(\ket{{\chi}_\text{out}}\bra{{\chi}_\text{out}}\right).\label{rho1}
\end{align}
In Eq.~\eqref{rho1}, $\ket{{\chi}_\text{out}}$ is given by Eq.~\eqref{chi_out},
with $\ket{{\chi}_\text{in}}$ parametrized by
\begin{align}
\ket{{\chi}_\text{in}}&=\left[\cos\!\frac{\theta_1}{2},e^{i\varphi_1}\sin\!\frac{\theta_1}{2}\right]^T
\!\!\!\otimes\left[\cos\!\frac{\theta_2}{2},e^{i\varphi_2}\sin\!\frac{\theta_2}{2}\right]^T\!\!\!,\label{chi_in}
\end{align}
where the Bloch sphere representation is adopted for nuclear qubits.
The solid angle differentials in Eq.~\eqref{EP} are given by
 $\mathrm{d}\Omega_1=\sin\theta_1\mathrm{d}\theta_1\mathrm{d}\varphi_1$
and $\mathrm{d}\Omega_2=\sin\theta_2\mathrm{d}\theta_2\mathrm{d}\varphi_2$.
For later convenience, the entanglement power at $p=0$ is named as the residual entanglement power.

\subsection{Concurrence}
\label{Conc}

The concurrence is a physical measure that quantifies the amount of entanglement in a specific quantum state \cite{Hill:1997}.
Given the out spin state parametrized by
\begin{align}
\ket{\chi_\text{out}}=\alpha\ket{00}+\beta\ket{01}+\gamma\ket{10}+\delta\ket{11},
\end{align}
the corresponding concurrence is defined by
\begin{align}
\Delta(p,\theta)=2|\alpha\delta-\beta\gamma|.
\end{align}
Here, the coefficients $\alpha$, $\beta$, $\gamma$, and $\delta$ obey
$|\alpha|^2+|\beta|^2+|\gamma|^2+|\delta|^2=1$ and generally depend on $p$ and $\theta$
for neutron-proton scattering.
Since $|\alpha\delta-\beta\gamma|\leq|\alpha||\delta|+|\beta||\gamma|\leq\frac{1}{2}(|\alpha|^2+|\beta|^2+|\gamma|^2+|\delta|^2)=\frac{1}{2}$,
the concurrence satisfies the property $0\leq\Delta(p,\theta)\leq1$. 
It can be shown that $\Delta(p,\theta)=0$ and $\Delta(p,\theta)=1$ give rise to unentangled and maximally entangled states, respectively.
Moreover, it is pointed out that the entanglement power is actually proportional to the concurrence of $\ket{\chi_\text{out}}$ averaged over all the possible $\ket{\chi_\text{in}}$ \cite{Low:2021ufv}.

\section{Spin entanglement}



\subsection{Exact results}
\label{NR}

 \begin{figure}

\centering
  \includegraphics[width=0.75\linewidth]{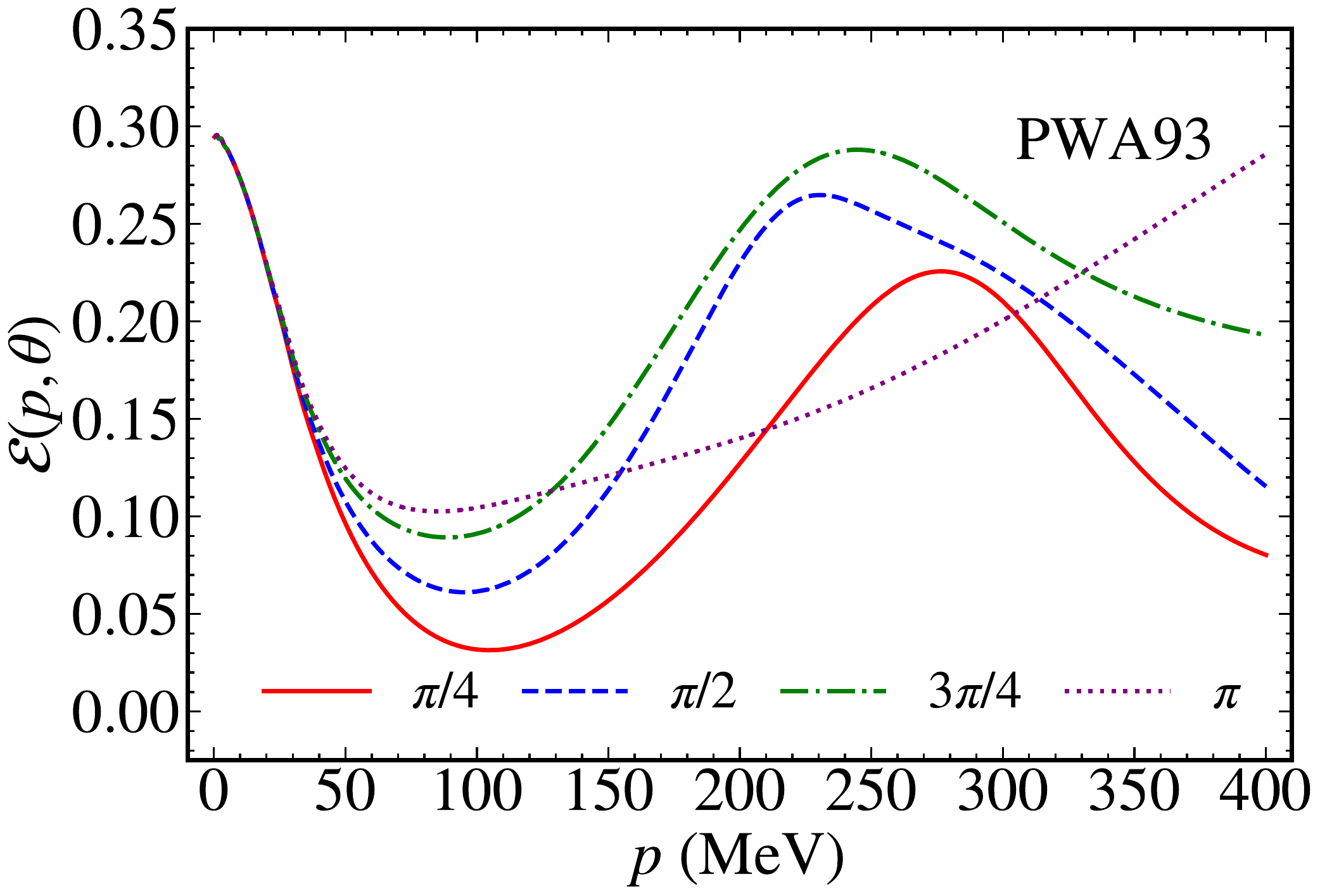}
  \caption{
  The entanglement power $\mathcal{E}(p,\theta)$ with respect to the scattering momentum $p$
  at four different relative angles $\theta=\frac{\pi}{4}$, $\frac{\pi}{2}$, $\frac{3\pi}{4}$, and $\pi$.
  The exact neutron-proton $S$-matrix inferred from the PWA93 model is used in numerical calculations.
  }

  \label{EP_p}
  
     
\end{figure}

 \begin{figure}

\centering
  \includegraphics[width=0.75\linewidth]{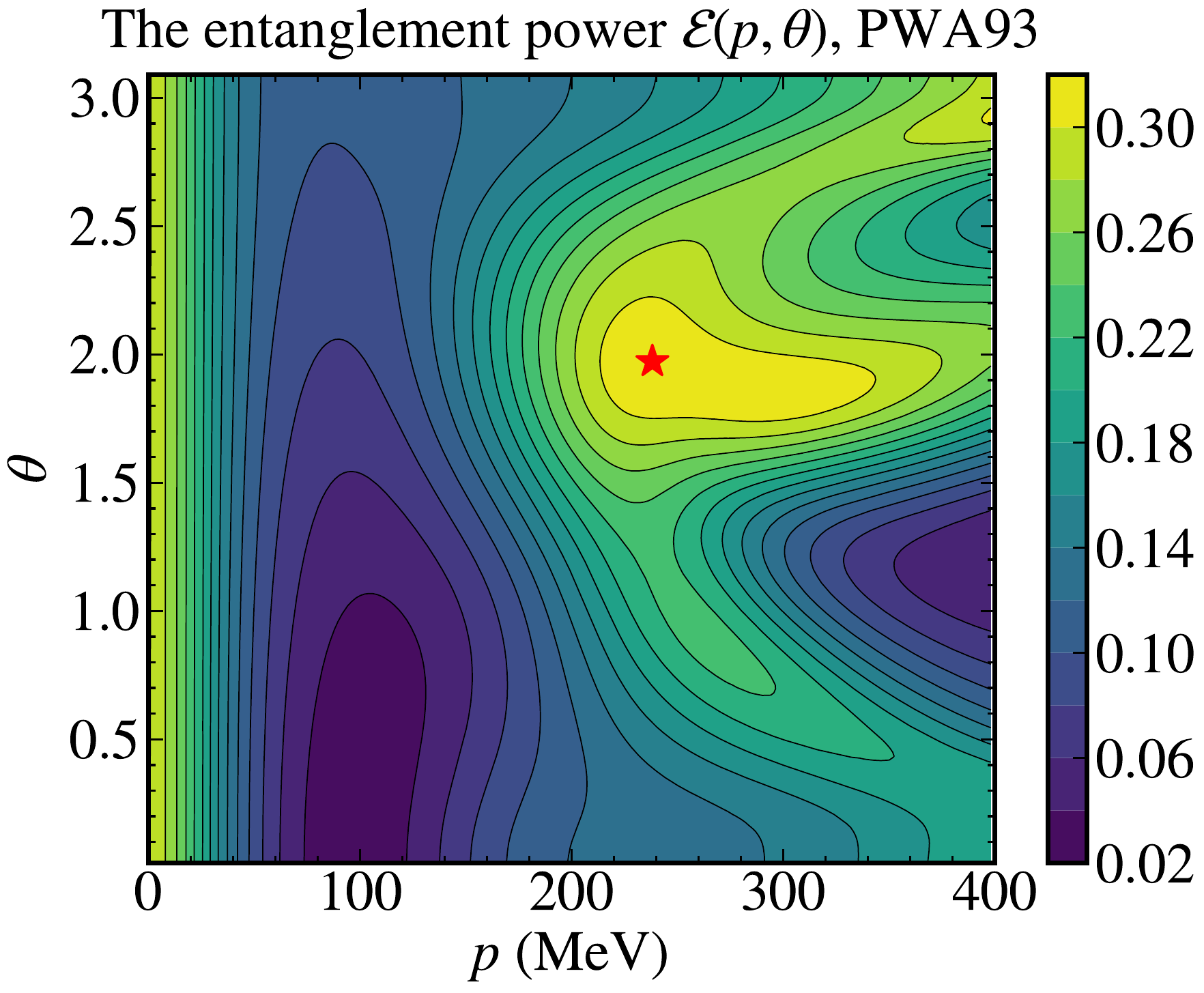}
  \caption{
  The contour plot of the entanglement power $\mathcal{E}(p,\theta)$ at $p\in[0,400]$ MeV and $\theta\in(0,\pi]$.
  The red star labels the maximal value of $\mathcal{E}(p,\theta)$ at $(p,\theta)\approx(238\ \text{MeV}, 113^\circ)$.
  The exact neutron-proton $S$-matrix inferred from the PWA93 model is used in numerical calculations.
}

  \label{EP_all}
  
     
\end{figure}

  \begin{figure*}

\centering
  \includegraphics[width=0.32\linewidth]{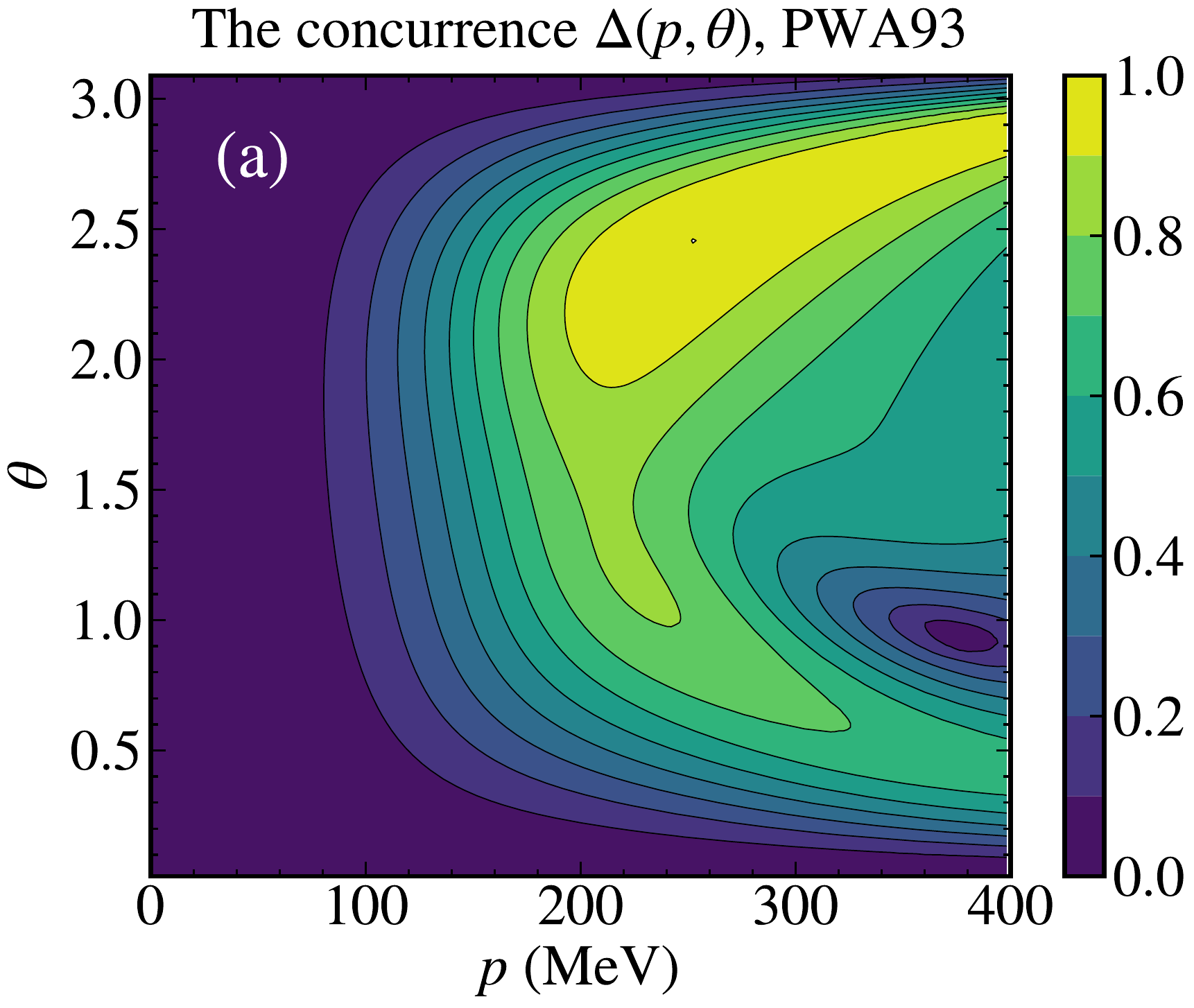}
  \includegraphics[width=0.32\linewidth]{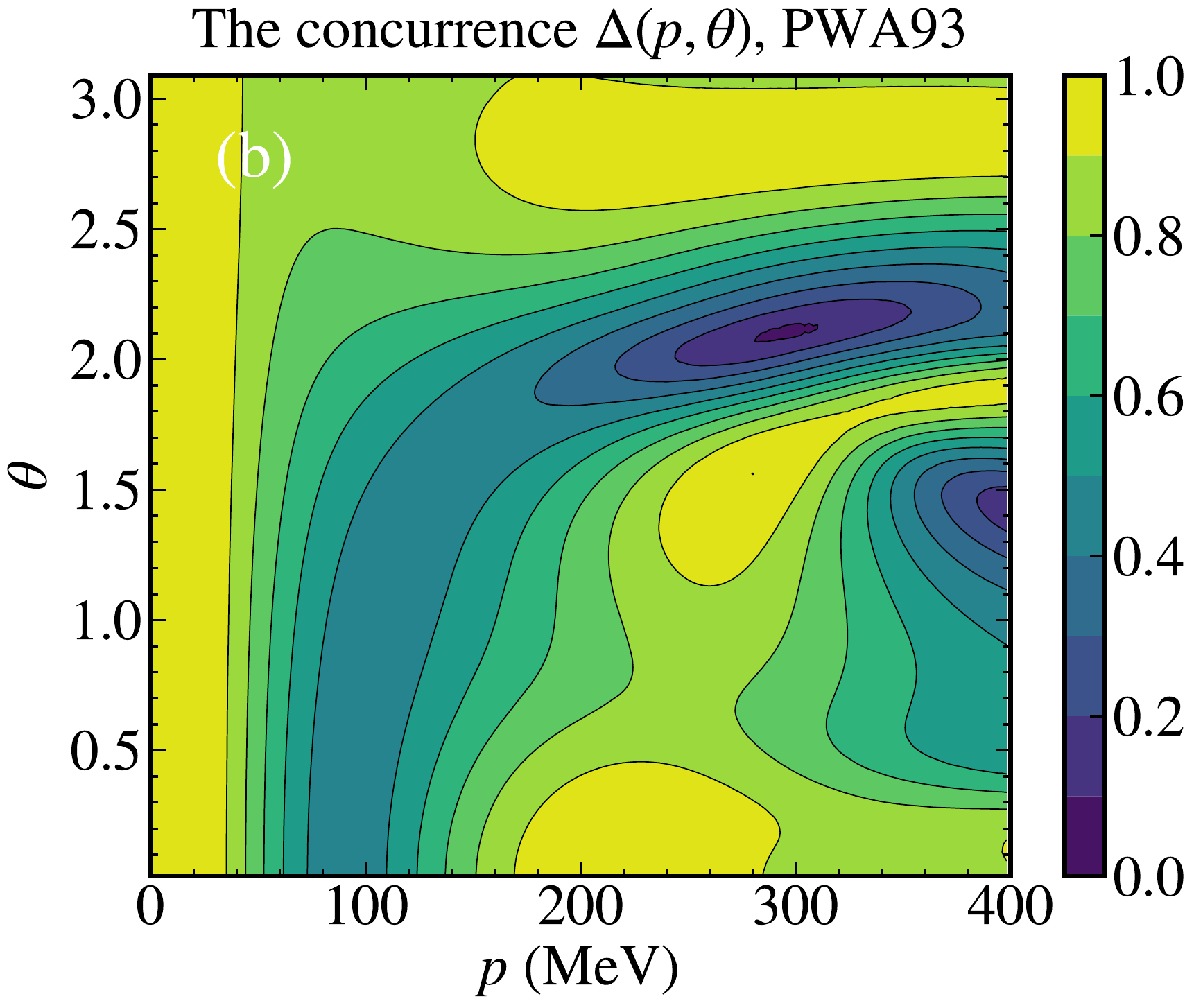}
  \includegraphics[width=0.32\linewidth]{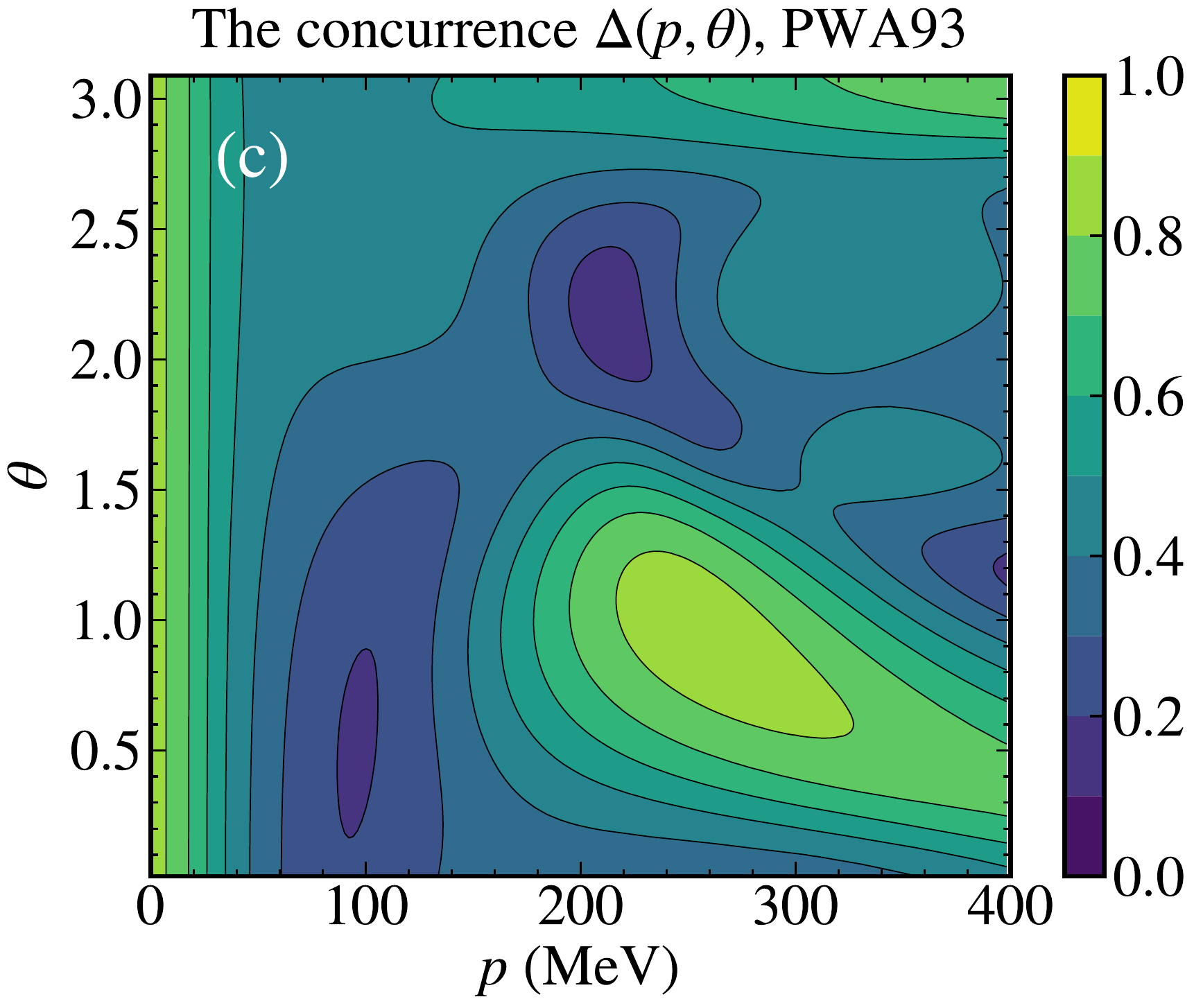}
  \includegraphics[width=0.32\linewidth]{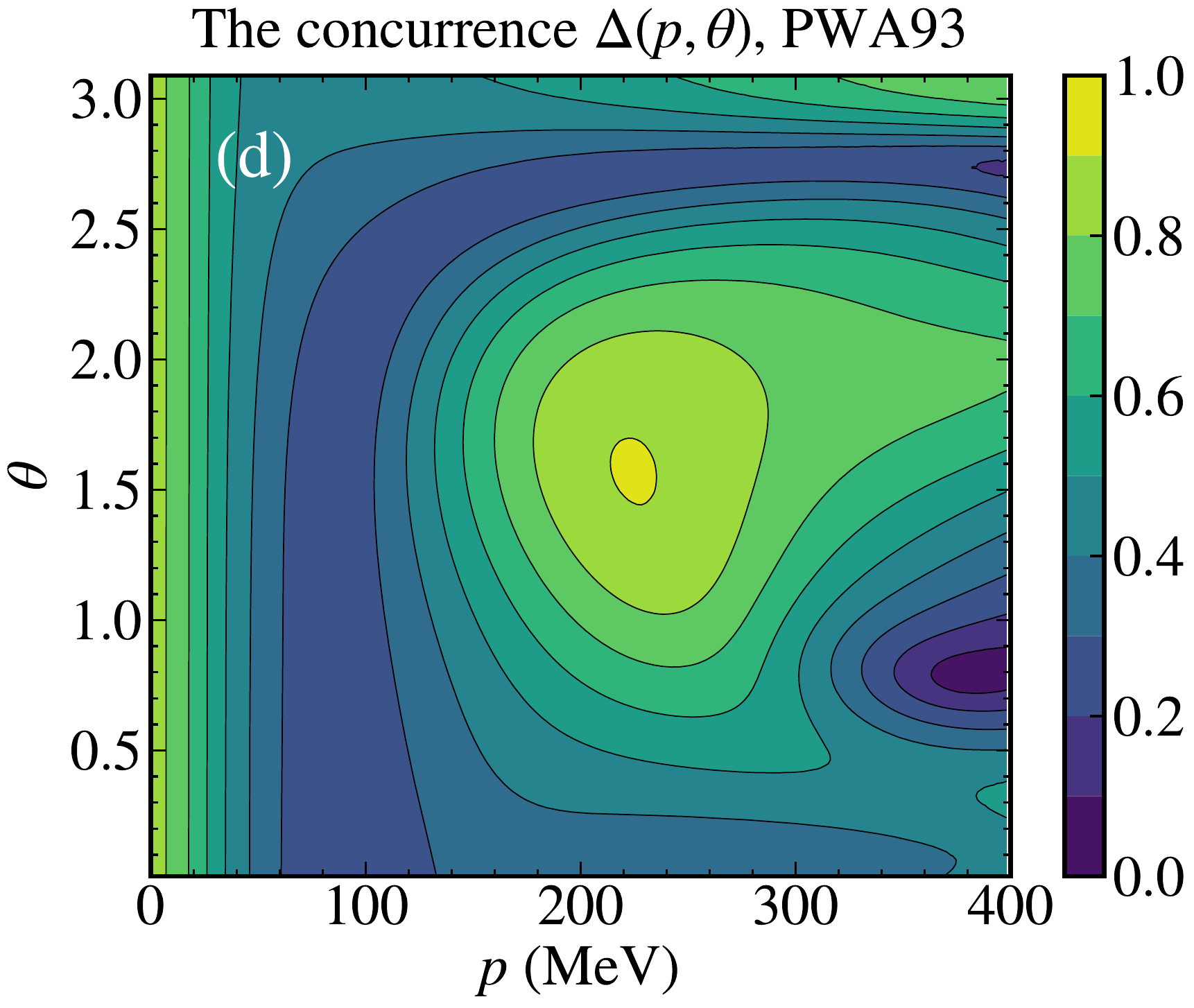}
  \includegraphics[width=0.32\linewidth]{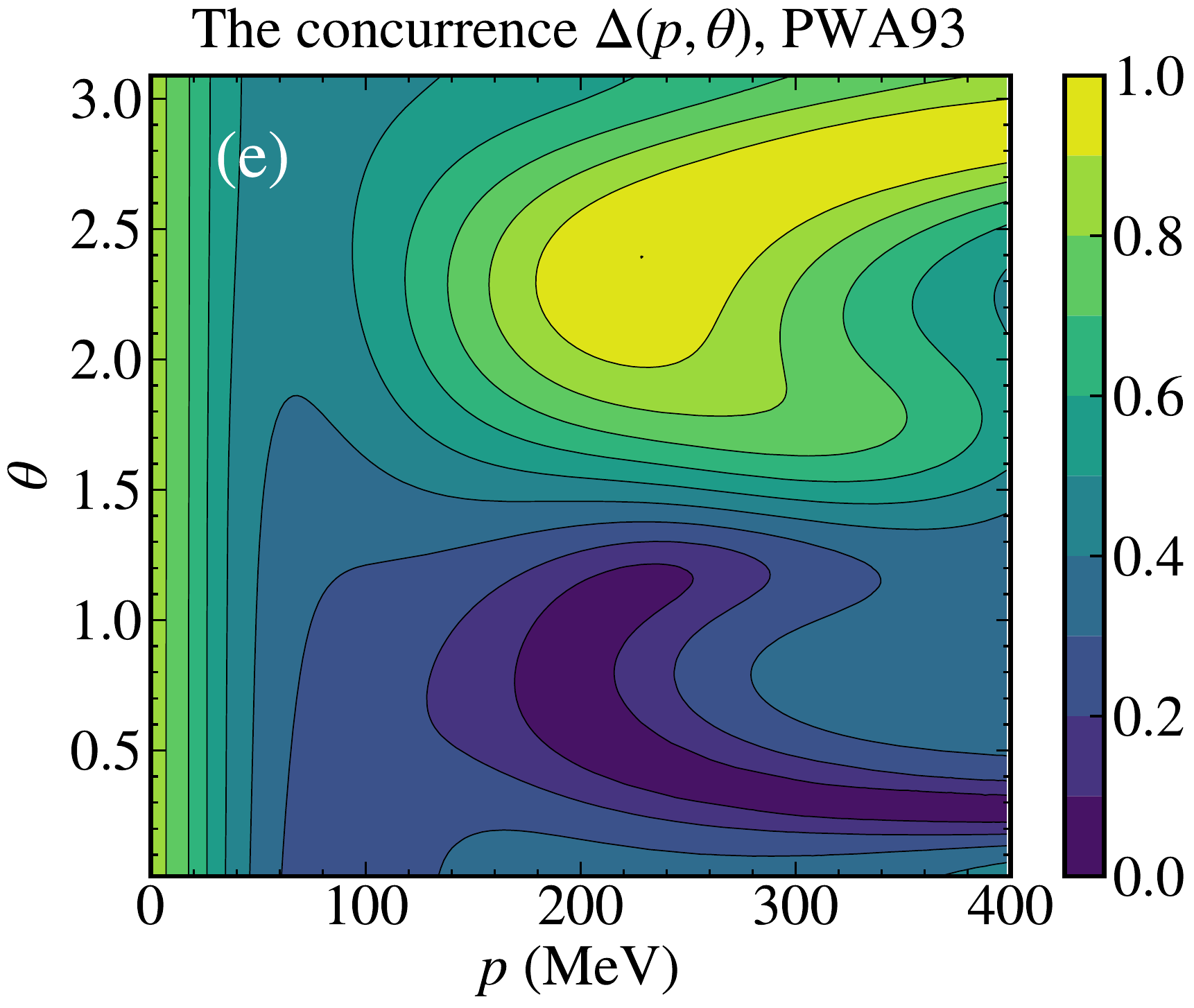}
  \includegraphics[width=0.32\linewidth]{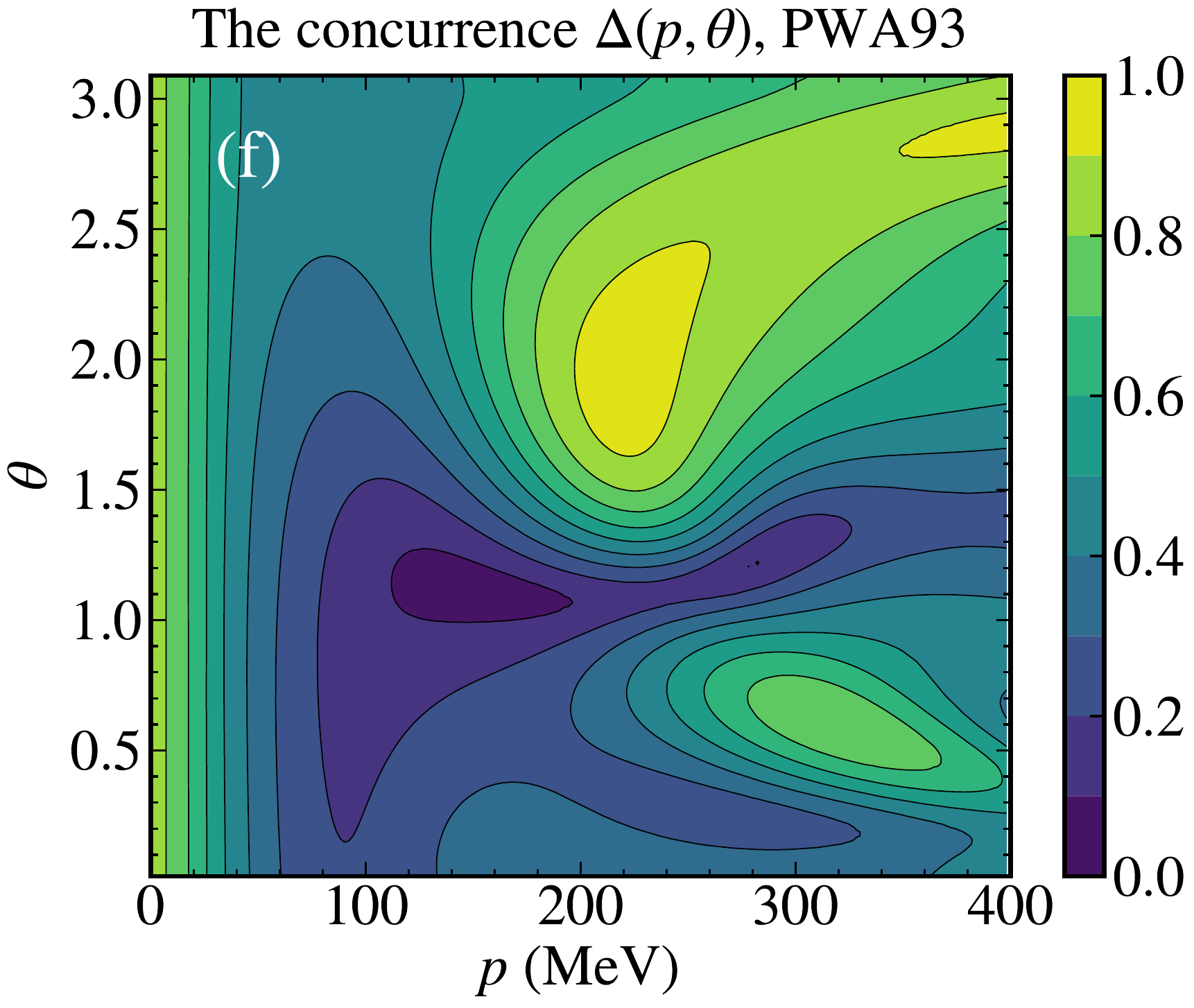}

  \caption{
  The contour plot of the concurrence $\Delta(p,\theta)$ at $p\in[0,400]$ MeV and $\theta\in(0,\pi]$
  for six specific choices of $\ket{\chi_\text{in}}$, including (a) $\ket{\uparrow\uparrow}$,
 (b) $\ket{\uparrow\downarrow}$, (c) $\ket{\uparrow}\otimes\frac{1}{\sqrt{2}}(\ket{\uparrow}+\ket{\downarrow})$, (d) $\ket{\uparrow}\otimes\frac{1}{\sqrt{2}}(\ket{\uparrow}-\ket{\downarrow})$, (e) $\ket{\uparrow}\otimes\frac{1}{\sqrt{2}}(\ket{\uparrow}+i\ket{\downarrow})$, and (f) $\ket{\uparrow}\otimes\frac{1}{\sqrt{2}}(\ket{\uparrow}-i\ket{\downarrow})$.
   The exact neutron-proton $S$-matrix inferred from the PWA93 model is used in numerical calculations.
}

  \label{C3}
  
     
\end{figure*}

In general, the entanglement power $\mathcal{E}(p,\theta)$ depends on the relative momentum $p$ and the scattering angle $\theta$. 
In Fig.~\ref{EP_p}, I study the $p$-dependence of $\mathcal{E}(p,\theta)$ at $\theta=\frac{\pi}{4}$, $\frac{\pi}{2}$, $\frac{3\pi}{4}$, and $\pi$.
The exact neutron-proton $S$-matrix is used in numerical calculations,
which is induced from the phase shift and amplitude data of the Nijmegen group in terms of Eqs.~\eqref{Saclay}--\eqref{Msmpm} \cite{NN-OnLine:2023}.
Explicitly, it is the PWA93 model that is used \cite{Stoks:1993}.
Here, I only consider relative momenta less than 400 MeV, as the impacts of internal structures of nucleons may not be negligible at higher momenta.
It is straightforward to see that $\mathcal{E}(p,\theta)$ is significantly enhanced as $p\to0$,
indicating that the entanglement generation capacity of nucleon-nucleon interaction is sizable at low energies.
At high energies, $\mathcal{E}(p,\theta)$ depends strongly on $\theta$.
At $\theta=\frac{\pi}{4}$, $\frac{\pi}{2}$, and $\frac{3\pi}{4}$,
$\mathcal{E}(p,\theta)$ first decreases at low momenta around $p\in[0,100]$ MeV, then grows at intermediate momenta around $p\in[100,250]$ MeV, and finally decrease at even higher momenta.
At $\theta=\pi$, 
$\mathcal{E}(p,\theta)$ behaves differently,
 first decreasing at low momenta around $p\in[0,100]$ MeV and then always increasing at higher momenta.
 Noticeably,  
 these four curves of $\mathcal{E}(p,\theta)$ coincide with each other approximately
 at low momenta less than 50 MeV.
 This property will be analyzed in detail in Sec.~\ref{LEL}.
  
 In Fig.~\ref{EP_all},
 the contour plot of the entanglement power $\mathcal{E}(p,\theta)$
 is shown for $p\in[0,400]$ MeV and $\theta\in(0,\pi]$.
 Here, $\theta=0$ is excluded as it corresponds to scattering in the forward direction, which is not considered in this Letter.
 At low momenta less than 50 MeV, the contours are approximately straight lines along the vertical direction,
which means that $\mathcal{E}(p,\theta)$ has a weak dependence on $\theta$ at these momenta.
At higher momenta, the contours of $\mathcal{E}(p,\theta)$ become complicated,
indicating a strong dependence on both $p$ and $\theta$.
The value of $\mathcal{E}(p,\theta)$ becomes maximal at $(p,\theta)\approx(238\ \text{MeV}, 113^\circ)$,
which means that $\ket{\chi_\text{out}}$ generated by neutron-proton scattering from an arbitrary $\ket{\chi_\text{in}}$
has the maximal spin entanglement \emph{on average} at this point.

In Fig.~\ref{C3}, the concurrence of $\ket{\chi_\text{out}}$ generated by specific $\ket{\chi_\text{in}}$ is shown.
Six choices are studied explicitly, including $\ket{\chi_\text{in}}=\ket{\uparrow\uparrow}$, $\ket{\uparrow\downarrow}$, $\ket{\uparrow}\otimes\frac{1}{\sqrt{2}}(\ket{\uparrow}+\ket{\downarrow})$, $\ket{\uparrow}\otimes\frac{1}{\sqrt{2}}(\ket{\uparrow}-\ket{\downarrow})$, $\ket{\uparrow}\otimes\frac{1}{\sqrt{2}}(\ket{\uparrow}+i\ket{\downarrow})$, and $\ket{\uparrow}\otimes\frac{1}{\sqrt{2}}(\ket{\uparrow}-i\ket{\downarrow})$.
As mentioned before, $\Delta(p,\theta)$ takes its values within $\Delta(p,\theta)\in[0,1]$, with $\Delta(p,\theta)=0$ corresponding to the completely unentangled $\ket{\chi_\text{out}}$
and $\Delta(p,\theta)=1$ corresponding to the maximally entangled $\ket{\chi_\text{out}}$.
It is straightforward to see that $\Delta(p,\theta)$ has a strong dependence on the explicit form of $\ket{\chi_\text{in}}$
and generally does not look like $\mathcal{E}(p,\theta)$ that averages $\Delta(p,\theta)$ over different choices of $\ket{\chi_\text{in}}$.
If experimentalists know initial spin configuration of neutron and proton exactly, 
they can manipulate spin entanglement properties of outgoing neutron-proton pairs in a controllable way.
If neutron-proton pairs with specific concurrence properties (e.g., $\Delta(p,\theta)=1$, maximally entangled) are wanted,
in principle,
experimentalists can produce them by
first working out a detailed contour plot of $\Delta(p,\theta)$ for experimentally accessible $(p,\theta)$,
then identifying the region on the $(p,\theta)$ plane that satisfies the required concurrence properties,
and finally collecting neutron-proton pairs at suitable $(p,\theta)$.

\subsection{Low-energy properties}
\label{LEL}

At low momenta $p\ll m_\pi$, neutron-proton scattering can be described by pionless effective field theory ($\cancel{\pi}$EFT) at leading order (LO),
whose Lagrangian is given by
\begin{align}
\mathcal{L}_\text{LO}&=N^\dagger\left(i\partial_t+\frac{\bm{\nabla}^2}{2M_N}\right)N \nonumber\\
&-\frac{C_S}{2}(N^\dagger N)(N^\dagger N)-\frac{C_T}{2}(N^\dagger\bm{\sigma}N)(N^\dagger\bm{\sigma}N),\label{PionlessEFT}
\end{align}
with $N$ being the nucleon field,
and $C_S$, $C_T$ being the low-energy constants (LECs).
The reduced neutron-proton $S$-matrix is given by
\begin{align}
&\!\!\hat{\bm{S}}(p)=\delta^2(\hat{p}'-\hat{p})\bm{1}_4+i\frac{p}{2\pi}\bm{M}(p),\label{S_LEL}\\
&\!\!\bm{M}(p)=\frac{1}{8ip}\left[(3e^{2i\delta_1}+e^{2i\delta_0}-4)\bm{1}_4+(e^{2i\delta_1}-e^{2i\delta_0})\bm{\sigma}\cdot\bm{\sigma}\right].\label{M_LEL}
\end{align}
Here, the ${}^1S_0$ (${}^3S_1$) phase shift $\delta_0$ ($\delta_1$) is related to the ${}^1S_0$ (${}^3S_1$) scattering length $a_0$ ($a_1$) by
\begin{align}
e^{2i\delta_{0,1}}=\frac{1+i\tan\delta_{0,1}}{1-i\tan\delta_{0,1}}=\frac{1-ipa_{0,1}}{1+ipa_{0,1}}.\label{ERE}
\end{align}
As shown by Eq.~\eqref{M_LEL},
the spin amplitude $\bm{M}(p)$ in $\cancel{\pi}$EFT at LO does not depend on the scattering angle $\theta$,
well consistent with the fact that $\mathcal{E}(p,\theta)$ has a weak dependence on $\theta$
at $p\ll m_\pi$ in Figs.~\ref{EP_p} and \ref{EP_all}.

When $p\to0$, the spin amplitude $\bm{M}(p)$ is given by
\begin{align}
\lim_{p\to0}\bm{M}&=
\left(
\begin{array}{cccc}
-a_1 & 0 & 0 & 0 \\
0 & -\frac{1}{2}(a_0+a_1) & \frac{1}{2}(a_0-a_1) & 0 \\
0 &  \frac{1}{2}(a_0-a_1) & -\frac{1}{2}(a_0+a_1) & 0 \\
0 & 0 & 0 & -a_1
\end{array}
\right),\\
&=-\frac{1}{2}(a_0+a_1)\bm{1}_4+\frac{1}{2}(a_0-a_1)\text{\footnotesize SWAP}.
\end{align}
In Fig.~\ref{EP_LEL}, the residual entanglement power $\mathcal{E}_0(a_0,a_1)$ at $p=0$  is plotted
as a function of $a_0$ and $a_1$,
excluding $(a_0,a_1)=(0,0)$ from the plot as $\mathcal{E}_0(a_0,a_1)$ is not defined at this point.
It is found that $\mathcal{E}_0(a_0,a_1)$ vanishes at $a_0=\pm a_1$ (i.e., taking its globally minimal value, white dashed lines).
$\lim_{p\to0}\bm{M}$ is proportional to the identity gate $\bm{1}_4$ at $a_0=a_1$,
and is proportional to the {\footnotesize SWAP} gate $\text{\footnotesize SWAP} = \frac{1}{2}(1+\bm{\sigma}\cdot\bm{\sigma})$ at $a_0=-a_1$.
In either case, it cannot generate spin entanglement from separable $\ket{\chi_\text{in}}$. 
On the other hand, $\mathcal{E}_0(a_0,a_1)$ is equal to 0.5 (0.1137) at $a_1=0$ ($a_0=0$) regardless of the value of $a_0\neq0$ ($a_1\neq0$) (red dashed lines),
corresponding to local maxima along the $a_0$ ($a_1$) direction.
The implications of these global and local extrema will be discussed in Sec.~\ref{EmSym}.
In the physical world, the ${}^1S_0$ and ${}^3S_1$ scattering lengths are found to be
$(a_0^\text{ph},a_1^\text{ph})\approx(-23.740,5.419)$ fm (red star), and the physical spin amplitude leads to $\mathcal{E}_0(a_0^\text{ph},a_1^\text{ph})\approx0.2953$.
However, it does not mean that the physical spin amplitude always produces entangled neutron-proton pairs.
For example, if $\ket{\chi_\text{in}}=\ket{\uparrow\uparrow}=[1,0,0,0]^T$,
$\ket{\chi_\text{out}}$ is equal to $\ket{\uparrow\uparrow}$ at $p=0$,
which turns out to be a separable state.
This explains why $\Delta(p,\theta)$ is much smaller than 1 at $p<100$ MeV in Fig.~\ref{C3}(a).

For completeness, I would like to mention a technical difference between my analysis and $S$-wave analyses carried out by Refs.~\cite{Beane:2018oxh,Low:2021ufv,Bai:2022hfv,Liu:2022grf,Bai:2023rkc}.
In Refs.~\cite{Beane:2018oxh,Low:2021ufv,Bai:2022hfv,Liu:2022grf,Bai:2023rkc}, it is the $S$-wave component of neutron-proton $S$-matrix $\hat{\bm{S}}(\bm{p}',\bm{p})=\delta^2(\hat{p}'-\hat{p})\bm{1}_4+\frac{ip}{2\pi}\bm{M}(\bm{p}',\bm{p})$ that is under consideration.
To be explicit, not only $\bm{M}(\bm{p}',\bm{p})$ but also $\delta^2(\hat{p}'-\hat{p})\bm{1}_4$ in the forward direction is decomposed into partial waves,
and only their $S$-wave components are kept for later processing. 
In this Letter, instead, it is low-energy neutron-proton $S$-matrix in the nonforward direction that is under consideration,
which, as shown by Eq.~\eqref{S_LEL}, equals the $S$-wave component of the spin amplitude given by Eq.~\eqref{M_LEL}.
This explains the difference between my results and those in Refs.~\cite{Beane:2018oxh,Low:2021ufv,Bai:2022hfv,Liu:2022grf,Bai:2023rkc}.

\subsection{Emergent symmetries}
\label{EmSym}

As mentioned before, nucleon-nucleon scattering at $p\ll m_\pi$ can be described by $\cancel{\pi}$EFT,
which has rich symmetries at LO.
If $C_T=0$ (equivalently, $a_0=a_1$),
$\cancel{\pi}$EFT is equipped with Wigner SU(4) symmetry at LO \cite{Wigner:1936dx},
as the spin-spin vertex vanishes in Eq.~\eqref{PionlessEFT}.
If $(|a_0|,|a_1|)=(0,0)$, $(0,\infty)$, $(\infty,0)$, and $(\infty,\infty)$,
$\cancel{\pi}$EFT is equipped with Schr\"odinger symmetry at LO, which is also known as nonrelativistic conformal symmetry \cite{Mehen:1999qs,Mehen:1999nd}.
The case of $(a_0,a_1)=(0,0)$ will be excluded in the following discussions,
as there is no neutron-proton scattering in this case.

\begin{figure}

\centering
  \includegraphics[width=0.75\linewidth]{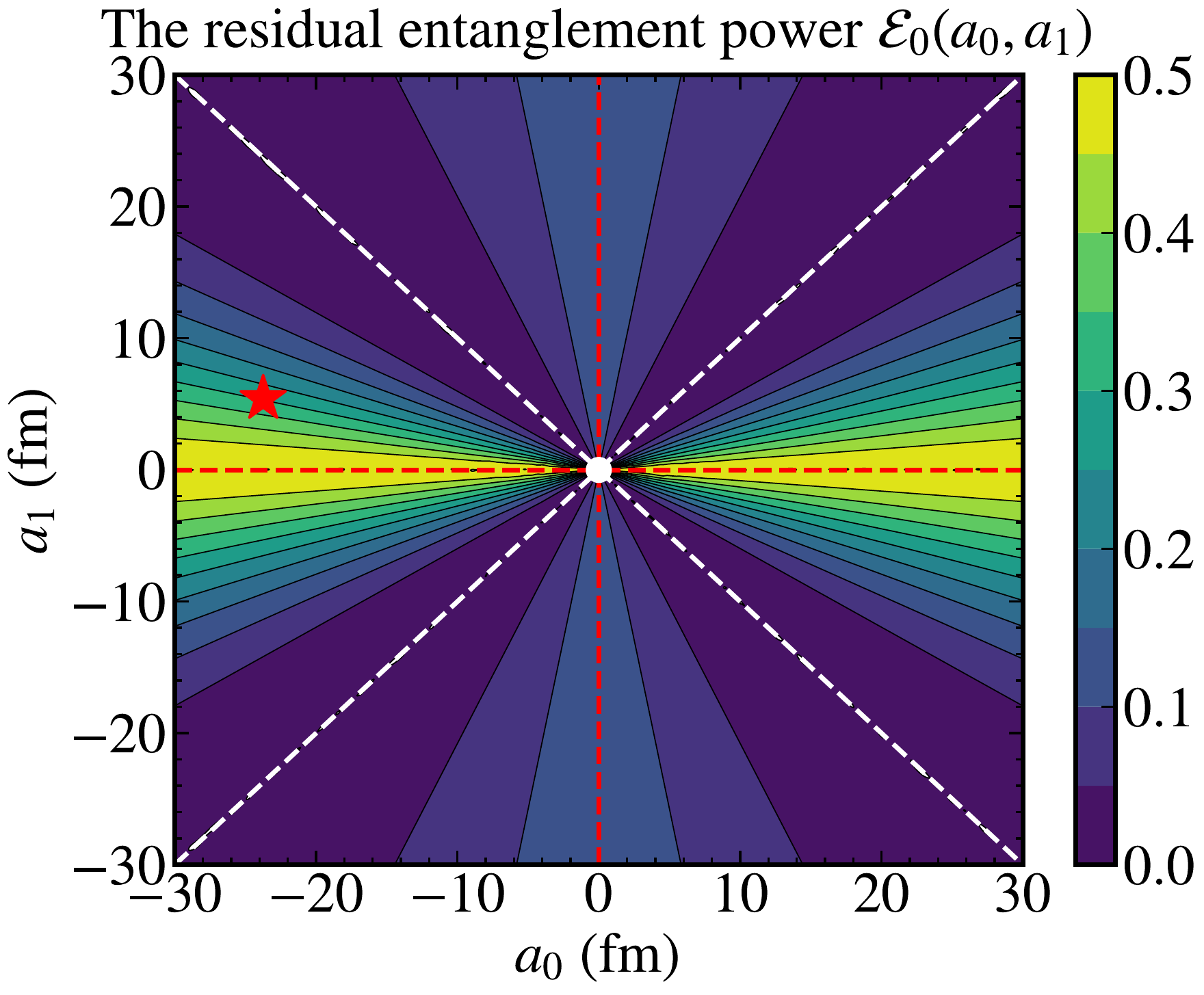}
  \caption{
  The contour plot for the residual entanglement power $\mathcal{E}_0(a_0,a_1)$ at $p=0$ as a function of the $^1S_0$ and $^3S_1$ scattering lengths $a_0$ and $a_1$,
  with $a_0,a_1\in[-30,30]$ fm. $\mathcal{E}_0(a_0,a_1)$ is not defined at $(a_0,a_1)=(0,0)$ (the white dot), as $\lim_{p\to0}\bm{M}$ equals zero correspondingly and
  there is no neutron-proton scattering.   
  The red star labels the physical scattering lengths at $(a_0,a_1)\approx(-23.740,5.419)$ fm.
  The white dashed lines label $a_1=a_0$ and $a_1=-a_0$, while the red dashed lines label $a_0=0$ and $a_1=0$.
}

  \label{EP_LEL}
  
     
\end{figure}

Several necessary conditions can be formulated in terms of spin entanglement for symmetry enhancement of $\cancel{\pi}$EFT at LO.
In Fig.~\ref{EP_LEL}, the residual entanglement power $\mathcal{E}_0(a_0,a_1)$ 
takes the globally minimal value equal to zero at $a_0=a_1$, where Wigner SU(4) symmetry emerges.
If $(a_0,a_1)=(\infty,\infty)$, $\mathcal{E}_0(a_0,a_1)$ vanishes as well, which can be regarded as a special example of $a_0=a_1$.
If $(a_0,a_1)=(\infty,-\infty)$ and $(-\infty,\infty)$, 
$\mathcal{E}_0(a_0,a_1)$  vanishes similarly, corresponding to special examples of $a_0=-a_1$ in Fig.~\ref{EP_LEL}. 
If $(|a_0|,|a_1|)=(\infty,0)$, $\mathcal{E}_0(a_0,a_1)$ takes the globally maximal value equal to 0.5.
If $(|a_0|,|a_1|)=(0,\infty)$, $\mathcal{E}_0(a_0,a_1)$ takes the locally maximal value equal to 0.1137.
In spite of different natures,
it is interesting to note that all these symmetry enhancement points are associated with global and local extrema of $\mathcal{E}_0(a_0,a_1)$.
This is a necessary condition for symmetry enhancement of $\cancel{\pi}$EFT at LO. 
It turns out that the same condition is also satisfied by
$a_0=-a_1=\text{finite number}$, $(a_0,a_1)=(\text{finite number},0)$, and $(a_0,a_1)=(0,\text{finite number})$,
where no symmetry enhancement is known.

Another necessary condition could be formulated by noticing that
the entanglement power $\mathcal{E}(p)$ remains unchanged at symmetry enhancement points with respect to relative momentum $p$.
In other words, symmetry enhancement points of $\cancel{\pi}$EFT at LO are associated with fixed points of $\mathcal{E}(p)$ in the variable of $p$.
I have made manifest only the dependence on $p$
to get consistent with notations in previous sections.
Explicitly,
if $a_0=a_1$, $\bm{M}=-\frac{a_0}{1+ia_0p}\bm{1}_4$, which depends on $p$ but does not generate spin entanglement due to its proportionality to the identity gate.
The $p$-dependent prefactor is then cancelled out in calculating $\mathcal{E}(p)$.
Similarly, if $(a_0,a_1)=(\infty,-\infty)$ and $(-\infty,\infty)$,
$\bm{M}=\frac{i}{p}\bm{1}_4$, which leads to $\mathcal{E}(p)=0$ at all momenta.
If $(|a_0|,|a_1|)=(\infty,0)$ and $(0,\infty)$, $\bm{M}=\frac{i}{2p}\{\{0,0,0,0\},\{0,1,-1,0\},\{0,-1,1,0\},\{0,0,0,0\}\}$
and $\frac{i}{p}\{\{1,0,0,0\},\{0,\frac{1}{2},\frac{1}{2},0\},\{0,\frac{1}{2},\frac{1}{2},0\},\{0,0,0,1\}\}$,
which turn out to be proportional to a constant matrix, with $p$-dependent prefactors cancelled out in calculating $\mathcal{E}(p)$.
Besides the symmetry enhancement points,
it is straightforward to check that $\mathcal{E}(p)$ at
$(a_0,a_1)=(\text{finite number},0)$ and $(0,\text{finite number})$
also remain invariant with respect to $p$.
Therefore, the condition of fixed points of $\mathcal{E}(p)$ is also not a sufficient condition
for symmetry enhancement in $\cancel{\pi}$EFT at LO.
In contrast to the first necessary condition formulated in term of global and local extrema of $\mathcal{E}_0(a_0,a_1)$,
the second necessary condition formulated in term of fixed points of $\mathcal{E}(p)$
is no longer satisfied by $\mathcal{E}(p)$ at $a_0=-a_1=\text{finite number}$.


In Ref.~\cite{Beane:2018oxh},
the connection between spin entanglement and symmetry enhancement has been studied in term of the entanglement power of the $S$-wave component of neutron-proton $S$-matrix (rather than the $S$-wave component of spin amplitude).
It is found that symmetry enhancement points in $\cancel{\pi}$EFT at LO are associated with zeros (i.e., global minima) of the entanglement power.
Clearly, the two necessary conditions proposed in this section are different from the connection found by Ref.~\cite{Beane:2018oxh}.

\section{Conclusions}
\label{Concl}

Nucleon-nucleon scattering is a fundamental process in low-energy nuclear physics.
In this Letter, I study spin entanglement properties of neutron-proton scattering with the exact $S$-matrix.
This goes beyond previous studies based on the $S$-wave component of $S$-matrix
and allows a more complete understanding of spin entanglement in neutron-proton scattering,
as well as its relation to neutron-proton interaction.
Within the framework of $\cancel{\pi}$EFT at LO, new connections are proposed for 
spin entanglement and symmetry enhancement of strong interactions at low energies.

Spin entanglement properties of neutron-proton scattering are not only valuable for understanding quantum entanglement properties of nuclear systems,
but also crucial for designing new kinds of nuclear scattering and reaction experiments by making use of quantum entanglement.
What new opportunities can be brought if entangled neutron-proton pairs are involved in nuclear scattering and reactions?
For instance,
suppose there is a neutron maximally entangled with a proton.
Is it possible to extract useful information of neutron-nucleus scattering by doing measurement on the proton?
At present, it is still too early to give any definite answer to questions of this kind.
This Letter takes the first step towards them
by laying the theoretical foundation for controllable production of entangled neutron-proton pairs,
the crucial building blocks of these processes.
As discussed in the main text, 
in principle,
this can be done by first working out the detailed dependence of spin entanglement on relative momentum, scattering angle, and initial spin configuration
and then collecting entangled neutron-proton pairs at suitable momenta and scattering angles for later processing.

%
%
%
%

\vskip 0.25cm

Note: This work was submitted for publication on May 4, 2023.
When it is under review, an independent work appears on arXiv, studying the same process and using very similar tools \cite{Miller:2023ujx}.
The results of Ref.~\cite{Miller:2023ujx} are well consistent with this work, except that it emphasizes slightly different aspects of spin entanglement in neutron-proton scattering.

\section*{Acknowledgments}
This work is supported by the Fundamental Research Funds for the Central Universities (Grant No.\ B230201022).

\end{document}